\documentclass{article}
\usepackage{spconf,amsmath,epsfig}
\usepackage{algorithm}
\usepackage{amssymb, algorithmic}
\usepackage{color,soul,comment}

\let\OLDthebibliography\thebibliography
\renewcommand\thebibliography[1]{
  \OLDthebibliography{#1}
  \setlength{\parskip}{0pt}
  \setlength{\itemsep}{0pt plus 0.3ex}
}

\begin{document}\sloppy

\def\x{{\mathbf x}}
\def\L{{\cal L}}

\title{Light Field Image Coding using Dual Discriminator Generative Adversarial Network and VVC Temporal Scalability}

%

\name{Nader Bakir$^{\ast \dag}$, Wassim Hamidouche$^{\ast}$, Sid Ahmed Fezza$^{\S}$, Khouloud Samrouth$^{\dag}$ and Olivier D\'eforges$^{\ast}$
}

\address{$^{\ast}$Univ. Rennes, INSA Rennes, CNRS, IETR - UMR 6164, Rennes, France \\
    $^{\dag}$Lebanese University, Tripoli, Lebanon\\
		$^{\S}$National Institute of Telecommunications and ICT, Oran, Algeria\\
whamidou@insa-rennes.fr}

\maketitle
\begin{abstract}
Light field technology represents a viable path for providing a high-quality VR content. However, such an imaging system generates a high amount of data leading to an urgent need for LF image compression solution. In this paper, we propose an efficient LF image coding scheme based on view synthesis. Instead of transmitting all the LF views, only some of them are coded and transmitted, while the remaining views are dropped. The transmitted views are coded using Versatile Video Coding (VVC) and used as reference views to synthesize the missing views at decoder side. The dropped views are generated using the efficient dual discriminator GAN model. The selection of reference/dropped views is performed using a rate distortion optimization based on the VVC temporal scalability. Experimental results show that the proposed method provides high coding performance and overcomes the state-of-the-art LF image compression solutions.
\end{abstract}
\begin{keywords}
Light Field, Deep Learning, D2GAN, VVC, Coding Structure, RDO.
\end{keywords}
\section{Introduction}
\label{sec:intro}
\begin{figure}[t!]
	\centerline{\includegraphics[width=0.39\textwidth]{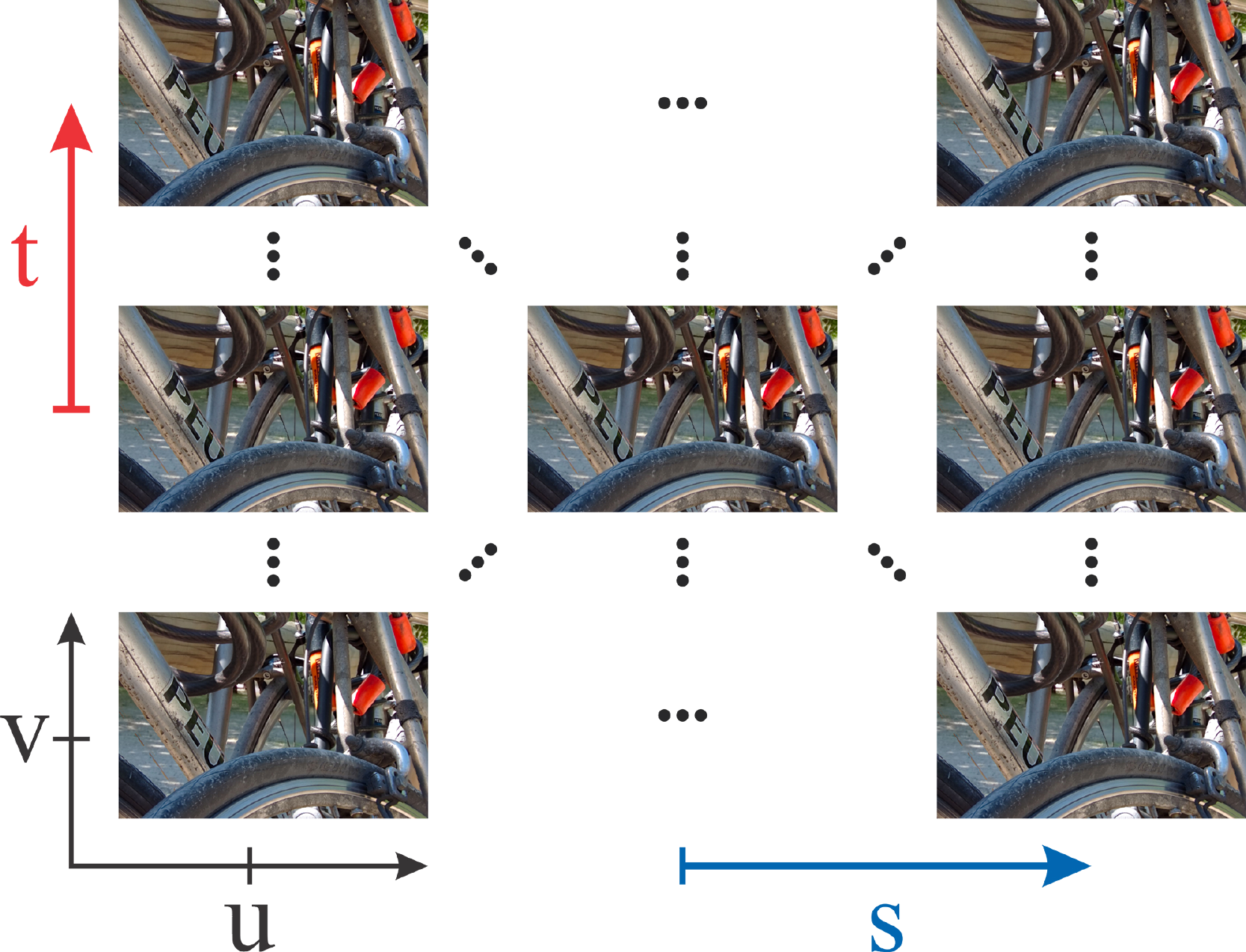}}
	\caption{Illustration of the Light Field image as an array of $u$, $v$ slices arranged in $s$, $t$.}
	\label{fig:subaperture}
\end{figure}

Light Field (LF) image can be captured and sampled by a plenoptic camera composed of an array of microlens such as \textit{Lytro} and \textit{Raytrix} cameras. LF imaging has many advantages over traditional 2D imaging systems, as its allows the user to change various camera settings after capture, thus providing more flexibility. Specifically, LF image captures both spatial and angular information enabling several multimedia services: multi-focus, multi-perspective, viewpoint rendering and even 6-Degree of Freedom (6-DoF) viewing~\cite{Guillemot:2017:LFIC, 8022901}. A light field can be described by a 4 dimensional function with 2 parallel plans $s,t$ and $u,v$ denoted by $L(u, \, v, \, s, \,t)$.
There are several ways to represent a LF image, including micro-image, epipolar image and sub-aperture image~\cite{Dansereau_light_2014}. The latter, illustrated in {\bf Fig.~\ref{fig:subaperture}}, is the the widely used representation. 

However, in sight of the huge amount of data involved by LF image, its processing, storage and transmission raise a real challenge that received increased research attention. In response, several solutions have been proposed to encode a LF image in sub-aperture representation. The straight forward coding approach organizes the LF views in a pseudo video sequence, which is then encoded with a classical 2D hybrid video encoder~\cite{7574674, 7805595, 8297145}. Another approach consists in encoding a spare set of views using a video encoder, while the rest of views are synthesized at the decoder side. The latter solution has been followed by several authors \cite{8297146,8712614,8574895}, for instance, linear approximation has been investigated in~\cite{8297146} to estimate the views at the decoder from neighbour views, while a combination of linear approximation and  convolutional neural network has been proposed in~\cite{8712614} to synthesize missing views at the decoder side. In the same way, Jia {\it et al.}~\cite{8574895} proposed to use the generative adversarial network to generate unsampled views. To enhance the coding efficiency, the authors proposed to encode and transmit the residual error between the generated uncoded views and their original versions. Jiang \textit{et al.}~\cite{8022889} proposed a coding method called Homography-based Low Rank Approximation. This method jointly optimizes multiple homographies that align the LF views and low rank approximation matrices. Hou {\it et al.}~\cite{Hou:2019:LFI:3347960.3347993} proposed a method that exploits the inter- and intra-view correlations effectively by characterizing its particular geometrical structure using both learning and advanced video coding techniques.

In this paper, we propose an efficient approach to encode the LF images, which consists in encoding a sparse set of views, and estimate the rest of views at the decoder side. In particular, the first set of selected reference views are coded with the next generation video coding standard called Versatile Video Coding (VVC). While the second set of views are either synthesized from the first decoded set of views using a Dual Discriminator Generative Adversarial Network (D2GAN) or decoded by a VVC decoder. The D2GAN have been trained with a large set of LF images coded at different distortions. The architecture offered by the D2GAN, composed by a generator and two discriminators, enables better training and thus synthesizes views with high visual quality. In addition, to increase the coding efficiency, a rate distortion optimization (RDO) is adopted to select which views should be encoded and transmitted and which ones should be dropped and synthesized at the decoder side.   


The remainder of this paper is organized as follows. Section~\ref{sec:related_works} describes the concepts of D2GAN and VVC. Then, in Section~\ref{sec:proposed_sol}, we describe the proposed LF image compression solution. Section~\ref{sec:RESULTSDISCUSSION} presents and discusses the experimental results. Finally, Section~\ref{sec:Conclusion} concludes this paper. 
\section{BACKGROUND}
\label{sec:related_works}
As mentioned in Section~\ref{sec:intro}, the proposed coding approach is based on  Dual Discriminator Generative Adversarial Network (D2GAN) and Versatile Video Coding (VVC) standard. In this section, we briefly introduce these two concepts.
\subsection{Dual Discriminator Generative Adversarial Nets}
\label{subsec:D2GAN}
Generative Adversarial Networks (GANs) are deep neural net architectures composed of two consecutive neural network models, namely generator $G$ and discriminator $D$. GAN enables to simultaneously train the two models: the generative model $G$ that captures the data distribution, and the discriminative model $D$ that estimates the probability that a sample came from the training data rather than from the generator $G$~\cite{goodfellow2014generative}. GAN has recently achieved great success in various fields, especially in fake video generation, super-resolution and objects detection~\cite{{8099502},{Bai_2018_ECCV}}.

Dual discriminator generative adversarial network (D2GAN), is a novel framework based on GAN, which uses two discriminators $D_{1}$ and $D_{2}$, where $D_{1}$ tries to assign high scores for real data, and $D_{2}$ tries to assign high scores for the fake data. This technique uses the two discriminators to minimize the Kullback-Leibler (KL) divergence and reverse KL between the generated image and the target image~\cite{NIPS2017_6860}.

Formally, $D_{1}$, $D_{2}$ and $G$ now play the following three player minimax optimization game
\begin{equation}
\label{eq:D2GAN}
\begin{split}
&\min_{G} \max_{D1,D2} \jmath~(G,D1,D2) = \alpha \, \mathbb{E}_{x \backsim P_{data}} [\log{D_{1}(x)}] \\
& + \mathbb{E}_{z \backsim P_{z}} [-D_{1}(G(z))] + \mathbb{E}_{x \backsim P_{data}} [-D_{2}(x)] \\
& + \beta \, \mathbb{E}_{z \backsim P_{z}} [\log{D_{2}(G(z))}], 
\end{split}
\end{equation} 
where $z$ is a noise vector, $\mathbb{E}$ represents expected value, $x$ is the real data, $P$ represents the probability distribution, $\alpha$~and~$\beta$ are two hyper-parameters (0  $< \alpha$, $\beta \leq$ 1) to stabilize the learning of the model and control the effect of KL and reverse KL divergences on the optimization problem~\cite{NIPS2017_6860}.

\begin{figure*}[t]
	\centerline{\includegraphics[width=1\textwidth]{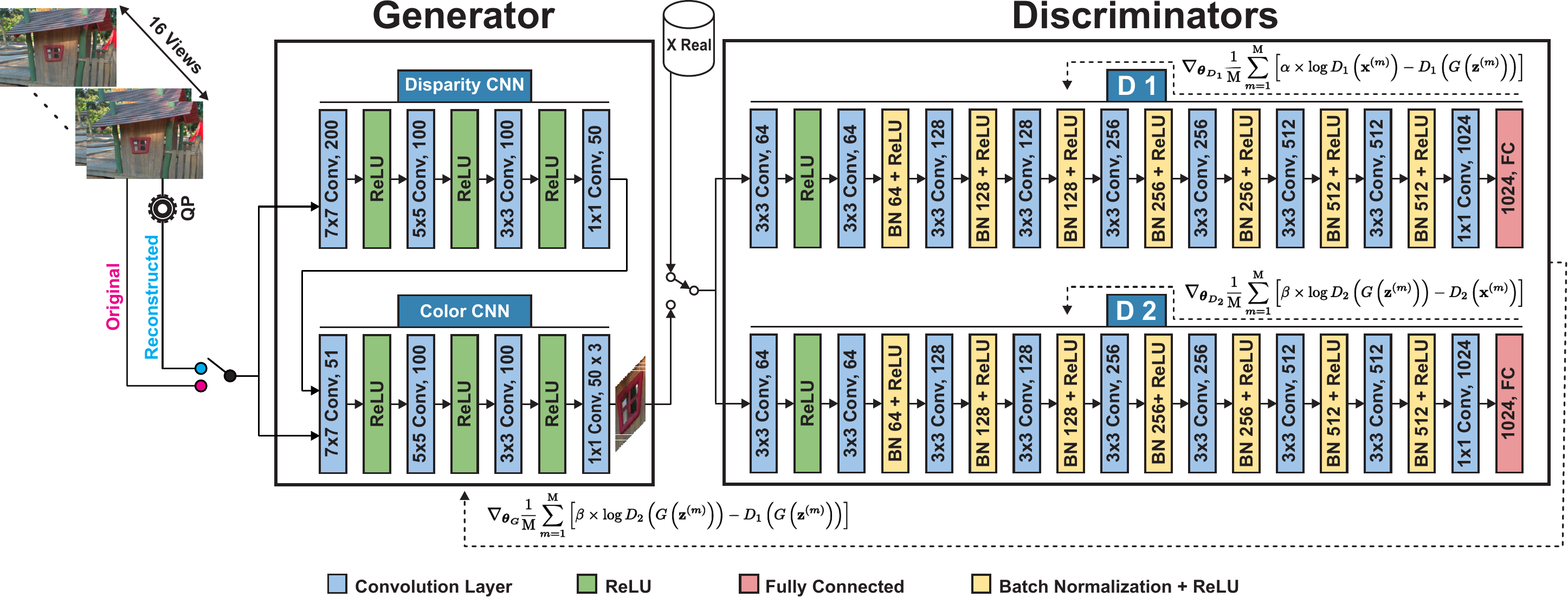}}
	\caption{Dual discriminator generative adversarial networks (D2GAN) architecture.}
	\label{fig:D2GANArchitecture}
\end{figure*}
More specifically, with a batch of $M$ noise samples $z^{(1)}, z^{(2)}, ..., z^{(M)}$ given as inputs, the generator generates $M$ artificial samples, and this process is defined as $G(z^{(i)})$. While, $x^{(1)}, x^{(2)}, ..., x^{(M)}$ represents a batch of $M$ real data samples.

Three cost functions defined in (\ref{eq:CostD1}), (\ref{eq:CostD2}) and (\ref{eq:CostG}) are computed to obtain the error that should be transmitted respectively to $D_{1}$, $D_{2}$ and $G$ for their backward weights updating, as shown in {\bf Fig.~\ref{fig:D2GANArchitecture}} (dash lines).
\begin{equation}
\label{eq:CostD1}
\begin{split}
\nabla_{\theta_{D1}} \dfrac{1}{M} \sum_{m=1}^{M}  [ \alpha \, \log{D_{1}(x^{(m)})} -D_{1}(G(z^{(m)}))],
\end{split}
\end{equation} 
\begin{equation}
\label{eq:CostD2}
\begin{split}
\nabla_{\theta_{D2}} \dfrac{1}{M} \sum_{m=1}^{M}  [ \beta \, \log{D_{2}(G(z^{(m)}))} -D_{2}(x^{(m)})],
\end{split}
\end{equation}
\begin{equation}
\label{eq:CostG}
\begin{split}
\nabla_{\theta_{G}} \dfrac{1}{M} \sum_{m=1}^{M}  [ \beta \, \log{D_{2}(G(z^{(m)}))} -D_{1}(G(z^{(m)}))].
\end{split}
\end{equation} 

In this work, we use D2GAN to synthesize the dropped LF views, where the generator consists of two conventional neural networks (CNN)~\cite{Kalantari:2016:LVS:2980179.2980251}. The first CNN estimates the disparity and the second one generates the color image.
\subsection{Versatile Video Coding}
\label{subsec:VVC}
\begin{figure}[t]
	\centerline{\includegraphics[width=0.49\textwidth]{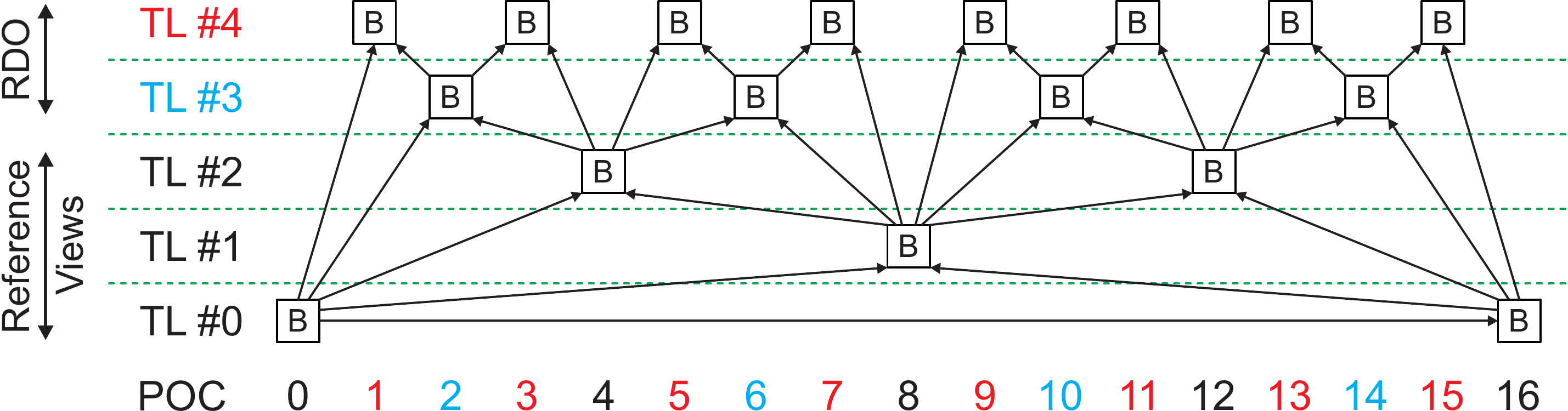}}
	\caption{Hierarchical prediction structure in VVC in Random Access (RA) coding configuration.}
	\label{fig:VVCGOP}
\end{figure}

Based on High Efficiency Video Coding (HEVC), the Joint Video Exploration Team (JVET) is developing a new video coding standard called Versatile Video Coding (VVC)~\cite{Wien2017}. VVC already enables a bitrate saving of 35\% to 40\% with respect to HEVC for the same visual quality~\cite{PCSVVC}. VVC introduces several new coding tools at different levels of the coding chain including frame partitioning, Intra/Inter predictions, transform, quantization, entropy coding and in-loop filters. For more details about the VVC coding tools the reader can refer to~\cite{8712638}. VVC supports by design the temporal scalability through the Random Access (RA) coding configuration. This latter, illustrated in {\bf Fig.~\ref{fig:VVCGOP}}, enables different temporal layers and each temporal layer uses as reference only frames from lower temporal resolution, \textit{i.e.}, lower layer. Therefore, frames of each temporal layer $t_i$ can be removed without impacting the decoding of frames of lower temporal resolution $t_j$ with $t_i>t_j$.

In the proposed coding approach, we exploit the concept of temporal resolution to drop views at the encoder without impacting the decoding process and thus performing the best rate distortion performance.        
\section{Proposed LF image compression method}
\label{sec:proposed_sol}
The idea behind the proposed coding method is, instead of transmitting all the LF views, to drop a sub-set of views at the encoder side and synthesize them at decoder side, thus considerably reducing the required bitrate for LF images. To efficiently achieve that, we exploit the temporal scalability of VVC and use the D2GAN model, all within a rate distortion optimization process (RDO).  

At the encoder side, first, LF sub-aperture views are organized into groups of 16 views that form Groups of Pictures (GOPs), as illustrated in {\bf Fig.~\ref{fig:VVCGOP}}. Next, in each GOP, the images of temporal levels 0, 1 and 2 are encoded using the VVC codec, which constitute the reference views used later in the synthesis process at the decoder side. Then, the images at the remaining levels 3 and 4 are either coded using the  VVC codec or dropped. In contrast to fix the number of dropped views, in our approach this is done adaptively on the basis of the proposed RDO process described in the Algorithm 1 and explaining in the following.

As illustrated in {\bf Fig.~\ref{fig:VVCGOP}}, we apply RDO process on the 3 consecutive frames, \textit{i.e.},  frame $i$ at level 4, frame $i+1$ at level 3 and frame $i+2$ at level 4. It should be noted that if one of the views at temporal level 4 (frame $i$ or $i+2$) must be encoded using VVC, then the frame $i+1$ at level 3 is also encoded using VVC, because this layer will be used as a reference for the frames at temporal level 4.
\begin{algorithm}[t!]
	\caption{RDO block based Lagrangian optimization}
	\begin{algorithmic}
		\REQUIRE $\mathcal{J} \leftarrow \{\ \forall\ m,\ \forall\ v \in TL\#[3\ or\ 4],\ \mathcal{J} = D + \lambda \, R \}$
         m: metod\ \{VVC,\ D2GAN\}
		\FORALL{ $v \in TL\#4$}
		
			\IF{$\mathcal{J}(VVC)\ <\ \mathcal{J}(D2GAN)$}
		       \STATE Encode $v$ by VVC
		        \STATE flag($v$) $\leftarrow$ False
		     \ELSE
	        	\STATE generate $v$ by D2GAN
		        \STATE flag($v$) $\leftarrow$ True
		    \ENDIF
		
		\ENDFOR
		\FORALL{ $v \in TL\#3$}
		
		\IF{$\mathcal{J}(VVC)\ <\ \mathcal{J}(D2GAN)$}
		\STATE Encode $v$ by VVC
		\STATE flag($v$) $\leftarrow$ False
		\ELSE[flag(previous($v$)) and flag(next($v$))]
		\STATE generate $v$ by D2GAN
		\STATE flag($v$) $\leftarrow$ True
		\ENDIF		
		\ENDFOR		
	\end{algorithmic}
\end{algorithm}

Main reasons behind only considering the 2 upper levels exclusively to the RDO block are, firstly, after an extensive study, we found that these levels occupy together around 28\% of the total bitrate. Second, the views at the upper levels are not used as references in the VVC coding scheme.

Thus, we proposed a RDO block deciding which views from the upper level can be encoded using VVC or dropped and synthesized using D2GAN. To reach this goal, the encoder computes the rate distortion (RD) cost function $J$ given by (\ref{eq:cost_function}) for both the VVC decoded view and the one synthesized by the D2GAN.
\begin{equation} \label{eq:cost_function}
\mathcal{J} = D + \lambda \, R
\end{equation}
where $\lambda$ is the Lagrangian multiplier, $D$ is the distortion and $R$ is the rate in bits per pixel (bpp). To set the Lagrangian multiplier ($\lambda$), we empirically determine its value by testing a large set of LF images. We found that the value of $0.1$ for $\lambda$ is optimal and for which the Lagrangian optimization is giving the best performance.

\begin{table}[t!]
	\begin{center}
		\label{table::D2GANConfigurations}
		\caption{The average coding gains in terms of BD-BR of D2GAN, trained with reconstructed views, in comparison with the anchor D2GAN training with original views.
		}		
			\footnotesize
			\begin{tabular}{|c|c|c|c|c|}
				\cline{1-5}
				\multicolumn{1}{|c|}{~}&
				\multicolumn{2}{c|}{wPSNR-based}&
				\multicolumn{2}{c|}{SSIM-based}\\
				\cline{2-5}
				\multicolumn{1}{|c|}{~}&
				\multicolumn{1}{c|}{BD-BR} &
				\multicolumn{1}{c|}{BD-PSNR}&
				\multicolumn{1}{c|}{BD-BR}&
				\multicolumn{1}{c|}{BD-SSIM}\\
				\cline{1-5}
			\hline
				\begin{tabular}[c]{@{}c@{}}vs. D2GAN \\ Reconstructed\end{tabular} & -11.0\% & 0.25 & -20.3\% & 0.013 \\
				\cline{1-5}
				\begin{tabular}[c]{@{}c@{}}vs. D2GAN \\ Recons. separately\end{tabular}  & \textbf{-16.6\%} & \textbf{0.39} & \textbf{-25.5\%} & \textbf{0.022}\\
				\cline{1-5}
				\hline
		\end{tabular}
	\end{center}
\end{table}

At the decoder side, the dropped views are synthesized using D2GAN block. As a reminder, the D2GAN  is composed of a generator $G$ and two discriminators $D_{1}$ and $D_{2}$. $G$ consists of two CNNs ~\cite{Kalantari:2016:LVS:2980179.2980251}, the first CNN estimates the disparity and the second one generates the color image. A set of features (mean and standard deviation) of a sparse set of views (16 views) are fed to the disparity CNN that estimates the disparity at an intermediate view, and then used it to warp (backward) all the input views to the intermediate view. The second color CNN uses all the warped images, derived from the first CNN, to predict the color and synthesizes the dropped views.
\begin{figure*}[t!]
		\includegraphics[width=0.33\textwidth]{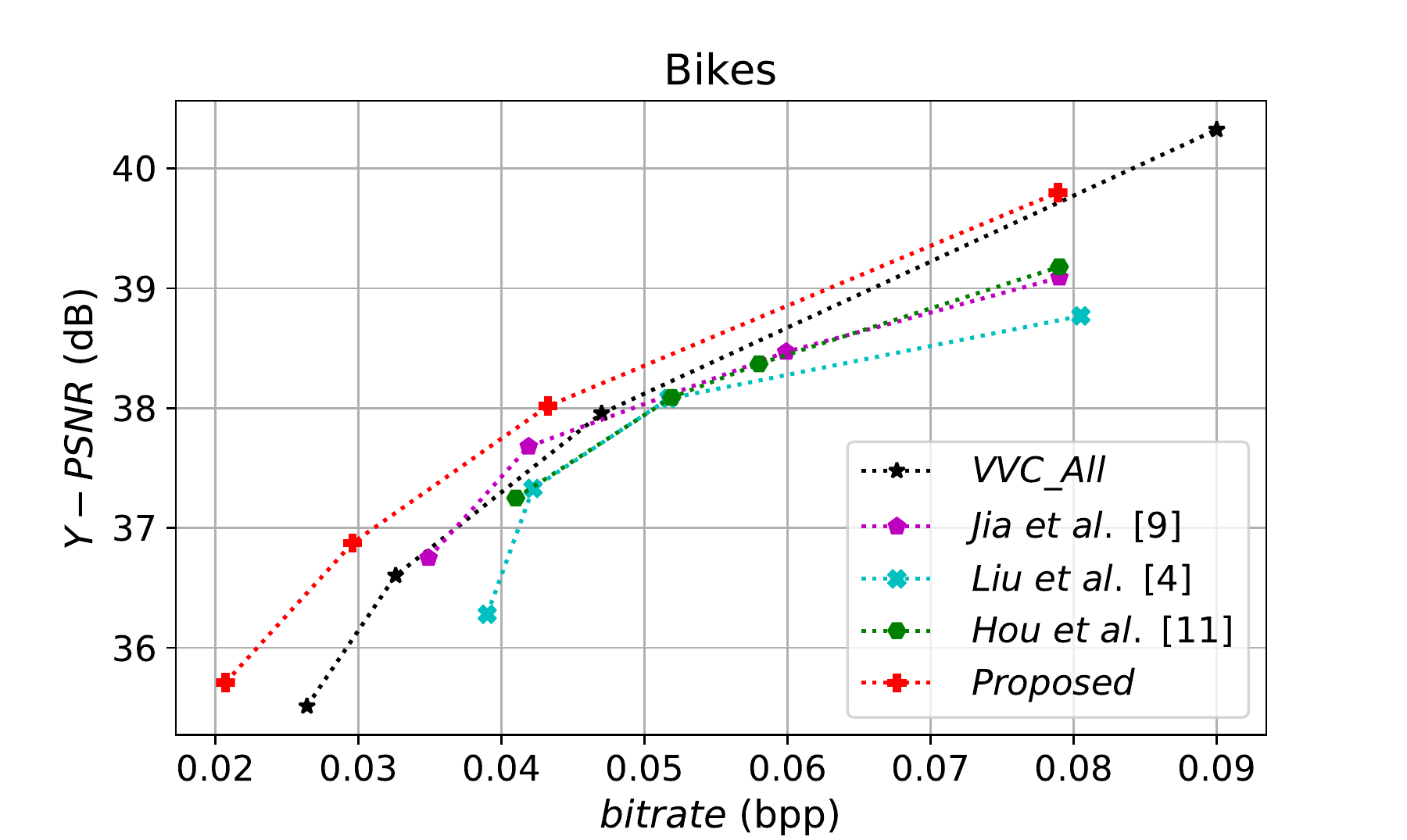}
		\includegraphics[width=0.33\textwidth]{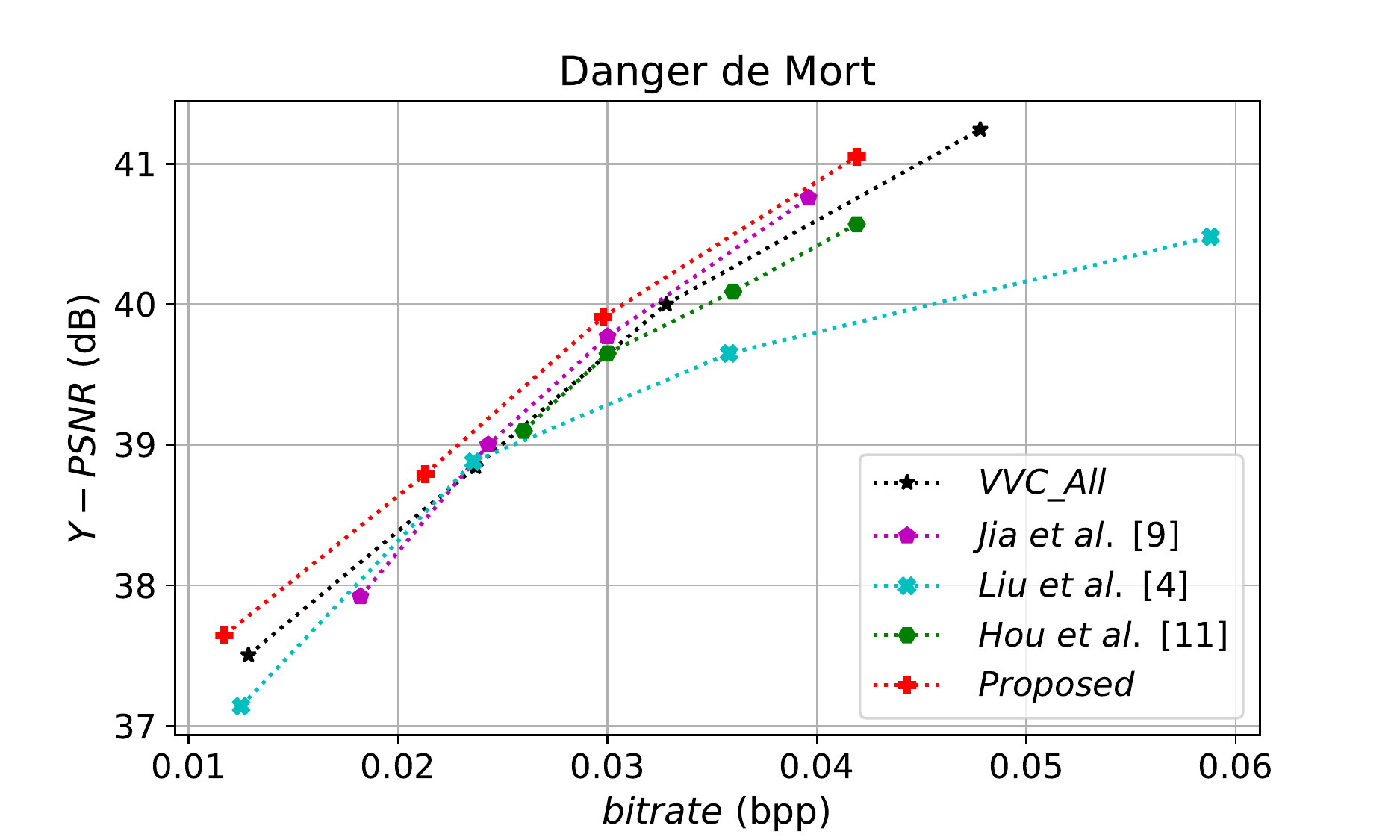}
		\includegraphics[width=0.33\textwidth]{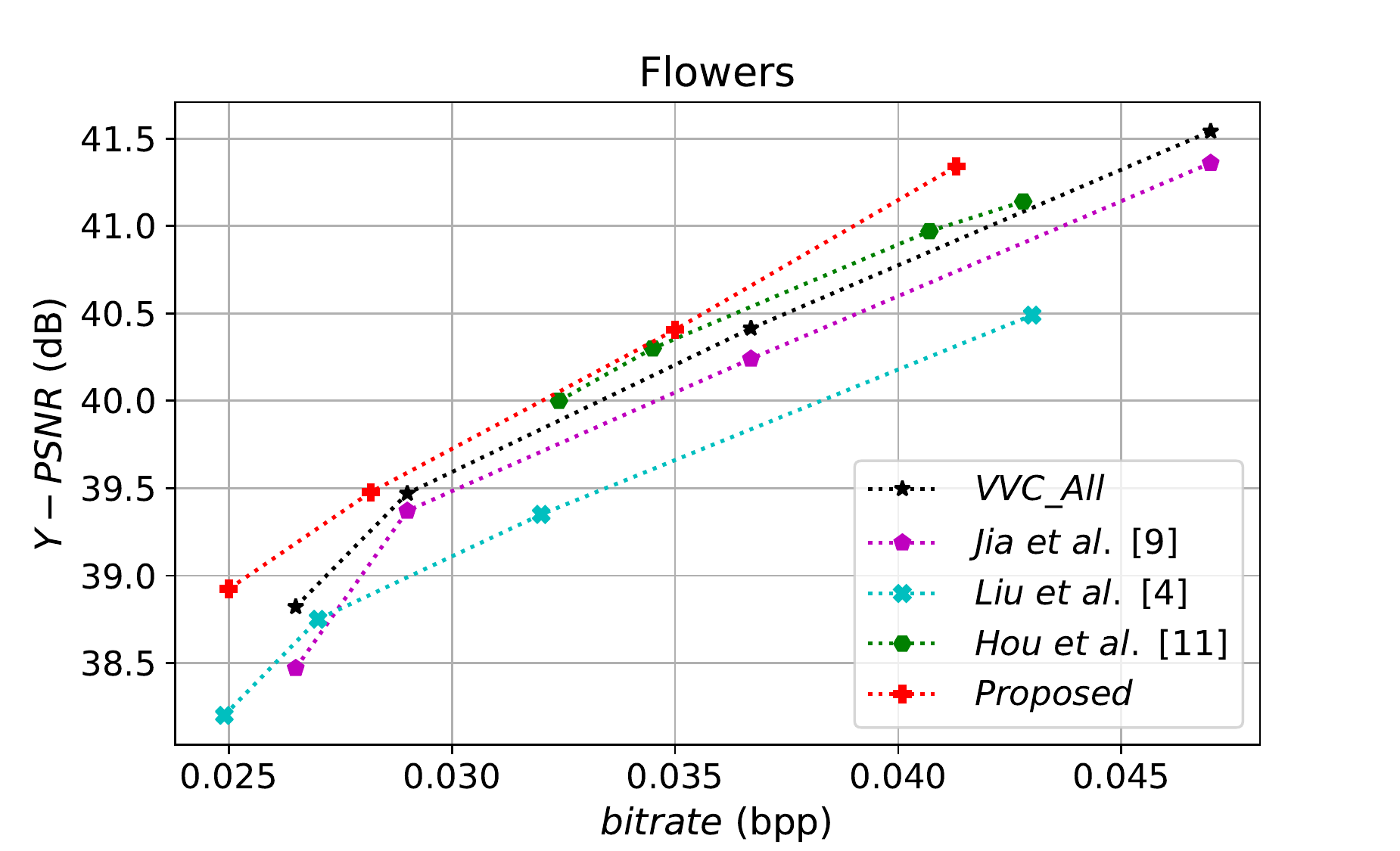}
		\includegraphics[width=0.33\textwidth]{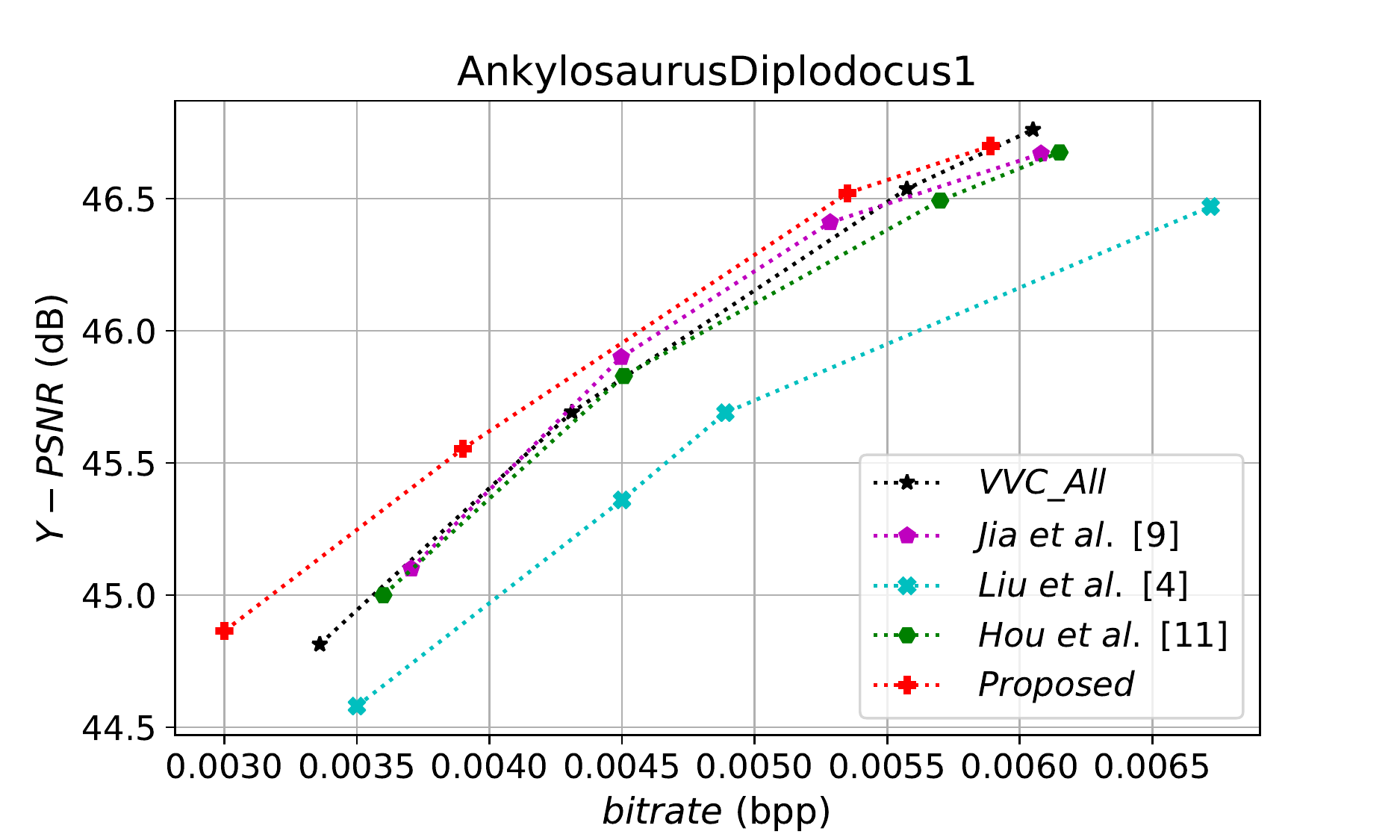}
		\includegraphics[width=0.33\textwidth]{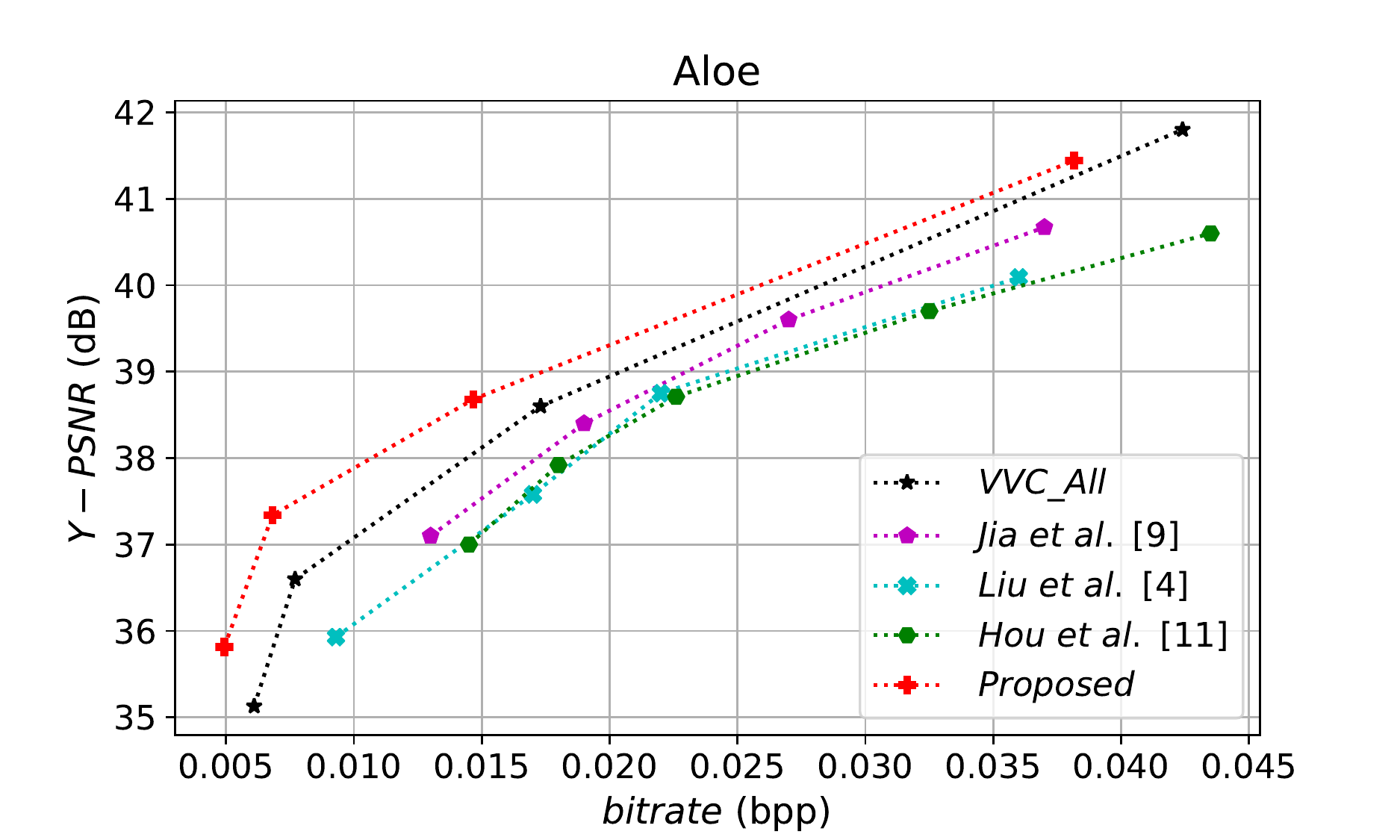}
	    \includegraphics[width=0.33\textwidth]{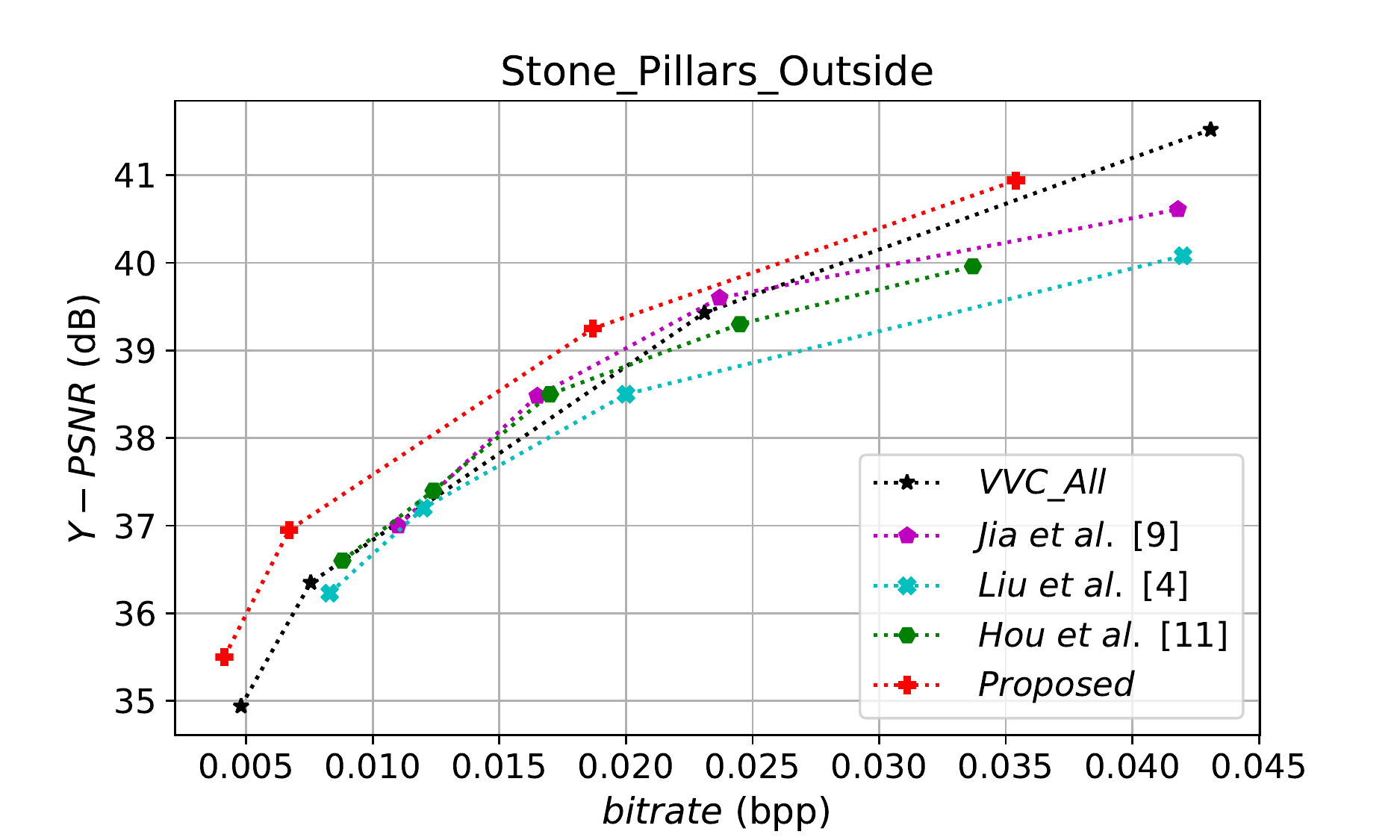}
	    \includegraphics[width=0.33\textwidth]{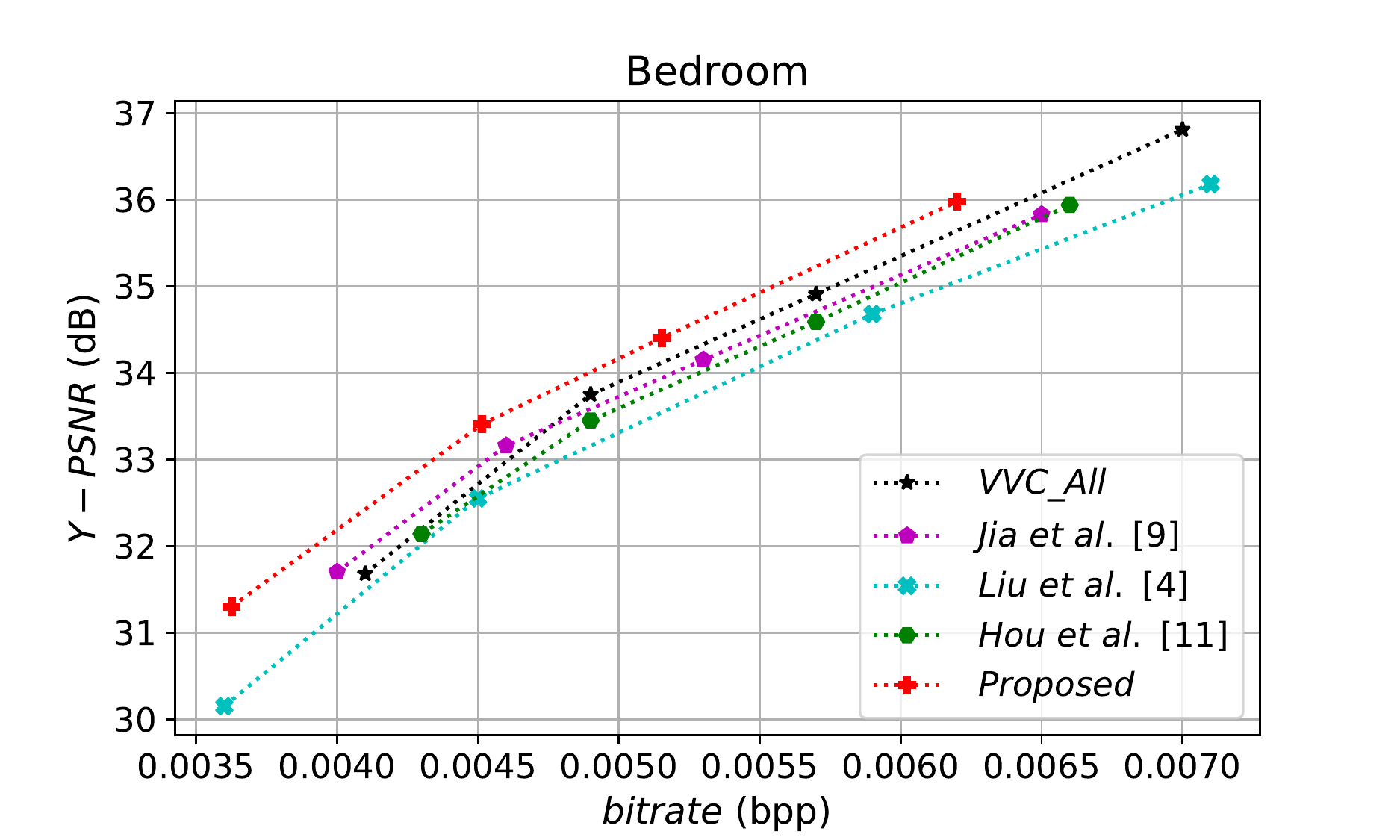}
	    \includegraphics[width=0.33\textwidth]{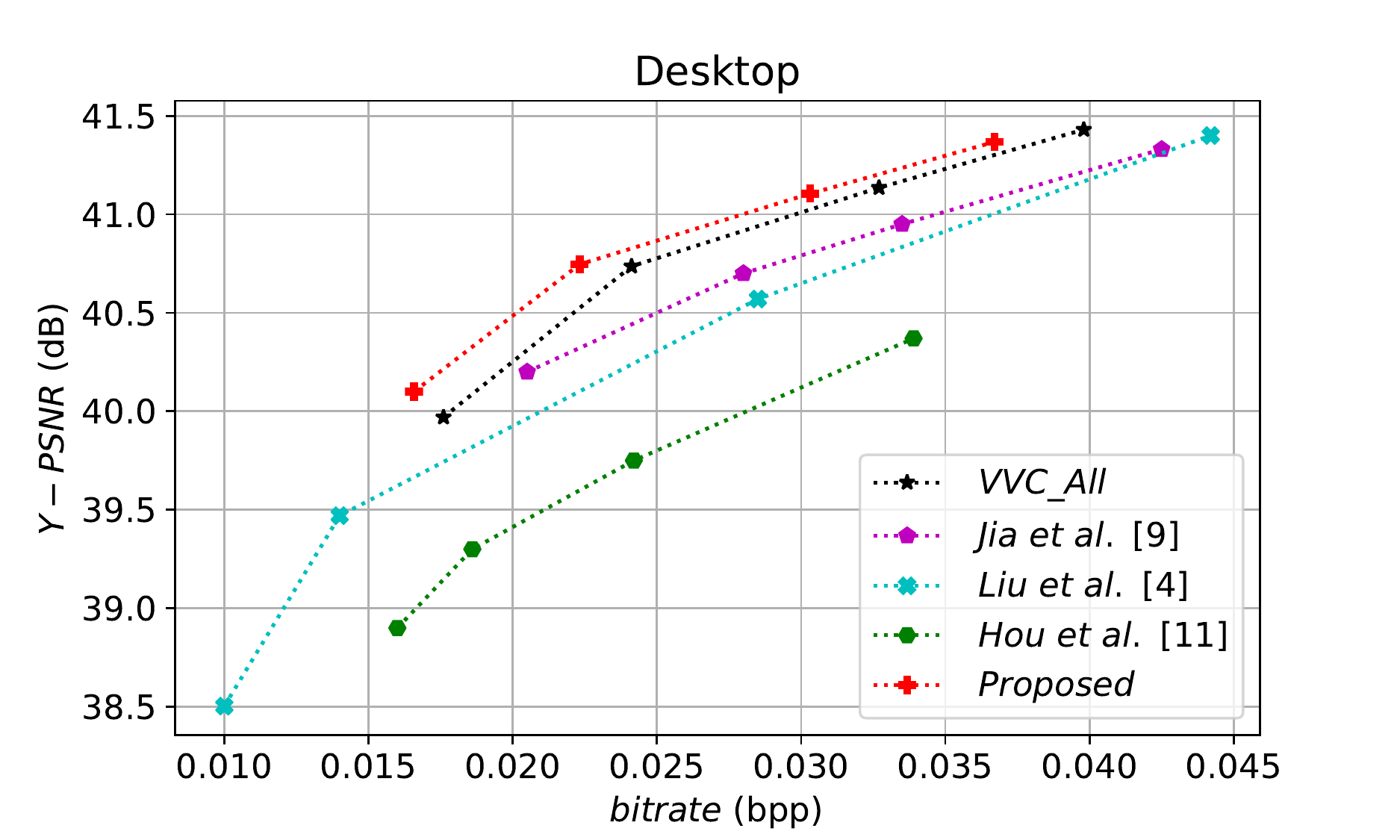}		
		\includegraphics[width=0.33\textwidth]{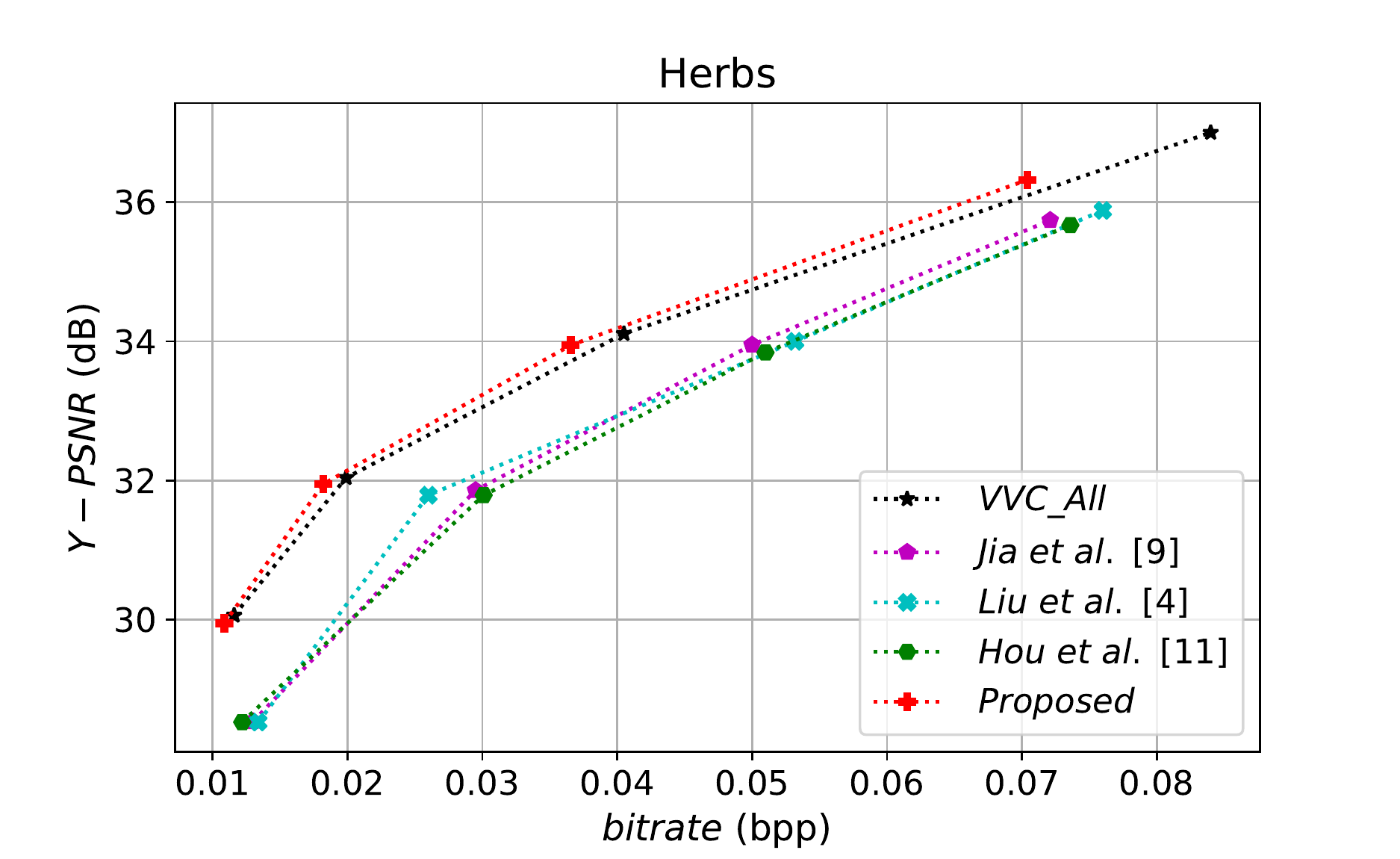}
	\caption{RD curves of the five considered solutions for the 9 LF images using four QP values.}
	\label{fig:results_psnr}
	\vspace{-3mm}
\end{figure*}
\begin{table*}[t!]
	\begin{center}
		\caption{BD-BR and BD-PSNR gains calculated against anchor method described in~\cite{7574674}.}
			\label{table:BDV}
			\footnotesize
			\begin{tabular}{|l|c|c|c|c|c|c|c|c|}
				\cline{1-9}
				\multicolumn{1}{|c|}{~}&
				\multicolumn{2}{c|}{VVC-All}&
				\multicolumn{2}{c|}{Jia \textit{et al.}~\cite{8574895}}&
				\multicolumn{2}{c|}{Hou \textit{et al.}~\cite{Hou:2019:LFI:3347960.3347993}}&
				\multicolumn{2}{c|}{Proposed}\\
				\cline{2-9}
				\multicolumn{1}{|c|}{Image} &
				BD-BR&BD-PSNR&BD-BR&BD-PSNR&BD-BR&BD-PSNR&BD-BR&BD-PSNR\\
				\hline
				\textit{Bikes} &--11.7\% &0.72  &--6.3\% &0.48 &--6.9\% &0.49  &\textbf{--22.4}\% &\textbf{0.96} \\			
				\textit{DangerDeMort} &--7.8\% &0.22 &--10.8\% &0.28 &--8.7\% &0.26 &\textbf{--16.5}\% &\textbf{0.40}\\
				\textit{Flowers} &--12.3\% &0.56 &--11.9\% &0.54 &--16.2\% &0.72 &\textbf{--16.6}\% &\textbf{0.74}\\				
				\textit{Ankylosaurus\_Dip1} &--13.2\% &0.44 &--14.9\% &--0.72 &--12.3\% &0.39 &\textbf{--18.0}\% &\textbf{0.57}\\				
				\textit{Aloe} &--26.4\% &0.85 &--9.1\% &0.31 &--2.46\% &--0.12 &\textbf{--42.3}\% &\textbf{1.23}\\				
				\textit{Stone\_pillars\_outside} &--18.3\% &0.61 &--15.1\% &0.52 &--11.9\% &0.28 &\textbf{--35.6}\% &\textbf{0.98}\\				
				\textit{Bedroom} &--5.3\% &0.46 &--4.0\% &0.32 &--2.3\% &0.18 &\textbf{--9.5}\% &\textbf{0.85}\\				
				\textit{Desktop} &--19.6\% &0.32 &--7.5\% &0.11 &44.1\% &--0.61 &\textbf{--26.3}\% &\textbf{0.45}\\				
				\textit{Herbs} &--26.0\% &1.14 &--4.4\% &--0.11 &6.9\% &--0.20 &\textbf{--29.8}\% &\textbf{1.32}\\				
				\cline{1-9}				
				Average  &--15.6\% &0.59 &--8.3\% &0.35  &--0.54\% &0.15 &\textbf{--24.1}\% &\textbf{0.83} \\							
				\hline
		\end{tabular}
	\end{center}
	\vspace{-5mm}
\end{table*}

Given that the generator $G$ and discriminators ($D1$ and $D2$) are CNN-based blocks, a training phase is required to fix respectively their parameters $\theta_{G}$, $\theta_{D1}$ and $\theta_{D2}$. Unlike GAN, in D2GAN, the scores returned by $G$ are values in $\mathbb{R}^{+}$ rather than probabilities in $\left[0,1\right]$. The discriminators and generator are alternatively updated using stochastic gradient ascent and descent, respectively. The backward propagation of errors (\textit{i.e.}, cost functions) is applied to update the discriminators and generator with mini-batch size equal to $M$, as shown in {\bf Fig.~\ref{fig:D2GANArchitecture}}.

For the training phase of D2GAN, 3 configurations were considered : 1) training with the original views , 2) training with reconstructed views at multiple distortion levels including the original views and 3) one training for each distortion level separately. We compared the three configurations, and the obtained results are given in {\bf Table~\ref{table::D2GANConfigurations}}. Based on these results, the third configuration, \textit{i.e.}, D2GAN reconstructed separately, outperforms the other configurations and hence we used it for the D2GAN training.


%
%
%
%
%
%

\section{Results and discussions}
\label{sec:RESULTSDISCUSSION}
\subsection{Experimental setup}
The proposed deep learning-based architecture described in the previous section was trained with 140 LF images, where 70 LF images are from EPFL dataset~\cite{Rerabek:218363}, 50 LF images are from Stanford Lytro LF image dataset~\cite{stanford2018} and 20 LF images are from HCI dataset~\cite{Honauer-etal-2017-ICCV}. Each sub-aperture view was split into patches of size 60$\times$60, thus resulting in more than 150,000 patches that were used in the training phase. For the testing phase, 9 LF images are selected, 6  LF images are from EPFL dataset~\cite{Rerabek:218363}, 1 LF image from Stanford Lytro LF dataset~\cite{stanford2018} and 2  LF images from HCI dataset~\cite{Honauer-etal-2017-ICCV}. Each of these LF images is composed of 8$\times$8 sub-aperture views. These views are rearranged in a pseudo sequence using spiral order scan and coded using VVC in random access (RA) coding configuration at 4 QP values of 18, 24, 28 and 32.

The training configuration of D2GAN was set as follows: we trained the generator $G$ and two discriminators ($D_{1}$ and $D_{2}$) with the ADAM optimizer by setting $\beta_{1} = 0.9$, $\beta_{2}= 0.999$, learning rate $= 0.0002$, batch-size of $10$ and kernel size of convolutional layers as depicted in Figure~\ref{fig:D2GANArchitecture}. The regularization coefficients of $D_{1}$ and $D_{2}$ were set as $\alpha= 0.2$ and $\beta= 0.2$, respectively. For the generator, we used input patch of 60$\times$60, stride of $16$, and output patch equal to $36\times36$ (reduced size is due to the convolutions).
\subsection{Evaluations}
\label{sec:evaluations}
We compared the proposed scheme with four state-of-the-art methods: 1) VVC-All that encodes all views with the VVC in RA coding configuration, 2) LF-GAN method proposed in~\cite{8574895}, where a sub-set of views are coded with HEVC, while the remaining views are generated by GAN and the residual error of views are transmitted to the decoder, 3) the method proposed in \cite{7574674} encodes the views as a pseudo-video sequence using specific order scan, 4) the method of Hou \textit{el al.} \cite{Hou:2019:LFI:3347960.3347993} that exploits the Inter- and Intra-views correlation to encode the views using HEVC. The latter method is considered as the anchor method.
\subsection{Results}
\label{sec:results}
The BD-BR~\cite{bjontegaard2008improvements} is a Peak Signal to Noise Ratio (PSNR) based metric. It is used in this paper to assess the gain of the proposed approach compared to the anchor solution. A negative BD-BR value refers to a bitrate reduction compared to the anchor method, while a positive value expresses a bitrate overhead.

R-D curves based on PSNR for the 9 LF images are provided in {\bf Fig.~\ref{fig:results_psnr}}. We can notice that for all considered images, the proposed coding method provides the highest performance for all bitrates. The previous conclusion is confirmed by {\bf Table~\ref{table:BDV}}, providing the Bj{\o}ntegaard results of the four coding solutions compared to the anchor one~\cite{7574674}. The proposed method achieved an average BD-BR gain of -24.1\% and BD-PSNR of 0.83 dB compared to the anchor method~\cite{7574674}.

The complexity of the proposed coding approach is also evaluated and compared to the other methods on both CPU and GPU platforms. The performance has been carried-out on an Intel core i9-7900X CPU running at 3.3GHz PC with 64 GB memory and a TITAN Xp NVDIA GPU. It is important to note that the GPU is only used when the D2GAN block is involved in the coding scheme. 

{ \bf Table~\ref{ComplexityEncoder}} gives the encoding and decoding run times in seconds. We can notice that the proposed solution requires almost the same complexity in the encoding for all QP compared to \cite{8574895} and \cite{Hou:2019:LFI:3347960.3347993} methods. The GPU enables to speedup the encoding part related to the D2GAN block. However, the decoder of the proposed solution is more complex than the other solutions due the D2GAN block.
%

\begin{table}[t!]
	\begin{center}
		\caption{Processing time in seconds of the four LF image coding methods.}
			\label{ComplexityEncoder}
		{
			\renewcommand{\baselinestretch}{1}\footnotesize
			\begin{tabular}{|c|c|c|c||c|c|}
				\cline{1-6}
				\multicolumn{1}{|c|}{~}&
				\multicolumn{5}{|c|}{Encoder}\\
				\cline{2-6}
				\multicolumn{1}{|c|}{QP}&
				\multicolumn{1}{c|}{VVC-All} &				
				\multicolumn{1}{c|}{Jia \textit{et al.}~\cite{8574895}}&
				\multicolumn{1}{c||}{Hou \textit{et al.}~\cite{Hou:2019:LFI:3347960.3347993}}&
				\multicolumn{2}{c|}{Our}\\
				\cline{2-6}
				\cline{2-6}
				\multicolumn{1}{|c|}{~}&
				\multicolumn{1}{c|}{CPU}&
				\multicolumn{1}{c|}{GPU}&
				\multicolumn{1}{c||}{CPU}&
				\multicolumn{1}{c|}{CPU}&
				\multicolumn{1}{c|}{GPU}\\
				\cline{1-6}
				\hline
				18 & \textbf{259} & 450 & 6028 & 559 & 449\\
				\cline{1-6}
				22 & \textbf{152} & 350 & 6028 & 452 & 342\\
				\cline{1-6}
				28 & \textbf{101} & 220 & 6028 & 401 & 291 \\
				\cline{1-6}
				34 & \textbf{66} & 142 & 6028 & 366 & 256 \\
\cline{1-6}
				Average & \textbf{66} & 291 & 6028 &  445 & 335 \\
				\hline\hline
				\multicolumn{1}{|c|}{~}&
				\multicolumn{5}{|c|}{Decoder}\\
					\hline
				Average & \textbf{4} & 53 & 583 & 124 & 94\\
				\hline
		\end{tabular}}
	\end{center} 
		\vspace{-6mm}
\end{table}

%
%
%
%
\section{Conclusion}
\label{sec:Conclusion}
In this paper, we have proposed a view synthesis based LF image compression approach. In the proposed coding scheme, a set of views are encoded using VCC, while the remaining views are dropped. The dropped views are synthesized using enhanced GAN-based approach known as D2GAN. The transmitted and dropped views are selected using RDO process. In addition, in order to avoid impacting the decoder with the dropped views, the latter are determined according to the temporal scalability of VCC. All these features allow reducing bitrate required by LF image, while providing views with high visual quality.

The experimental results showed the efficiency of our scheme, which achieved bitrate reduction of -24.1\% in terms of BD-BR and increased the visual quality by 0.83 dB in BD-PSNR with respect to the state-of-the-art solution.  
\bibliographystyle{IEEEbib}
\small
\bibliography{icme2020template}
\end{document}